\documentclass[12pt]{article}
\usepackage{amsfonts}
\usepackage{amsmath}
\usepackage{amssymb}
\usepackage{graphicx}%
\newtheorem{theorem}{Theorem}

\newtheorem{example}[theorem]{Example}

\begin{document}
\title{The Epidemics of Corruption}
\author{Ph.~Blanchard\thanks
{blanchard@physik.uni-bielefeld.de}
\and
A.~Krueger
\thanks{networks@andreaskrueger.de}
\and
T.~Krueger
\thanks{tkrueger@physik.uni-bielefeld.de}\\
University of Bielefeld, Faculty of Physics and BiBoS
\and
P.Martin
\thanks{peter.martin@schulz-berlin.de}\\
FU-Berlin, Department of Law}

\date{ }

 \maketitle

\begin{abstract}
We study corruption as a generalized epidemic process on the graph of social
relationships. The main difference to classical epidemic processes is the
strong nonlinear dependence of the transmission probability on the local
density of corruption and the mean field influence of the overall corruption
in the society. Network clustering and the degree-degree correlation play an
essential role in corruption dynamics. We discuss phase transitions, the
influence of the graph structure and the implications for epidemic control.
Structural and dynamical arguments are given why strongly hierarchically
organized societies like systems with dictatorial tendency are more vulnerable
to corruption than democracies. A similar type of modelling can be applied to
other social contagion spreading processes like opinion formation, doping
usage, social disorders or innovation dynamics.
\end{abstract}

\section{Introduction}

Corruption seems to be an unavoidable part of human social interaction,
prevalent in every society at any time since the very beginning of human
history till today. In sharp contrast to the high prevalence of corruption in
many countries and the rather large literature on political, social and
economical aspects of corruption there is only a small number of attempts to
model the dynamics of corruption in a mathematically quantified way. The
modelling approach in these few attempts essentially follows two paths. The
first is in the sense of microeconomics and incorporates game theoretic
aspects (for a recent model in this direction see the book by Steinr\"{u}cken
\cite{22} and the references therein) or rules for maximizing a certain
economically based profit functional (\cite{26}\cite{27}). Then a set of
differential equations for the evolution of the mean corruption is derived and
a stability analysis done on that basis. In these models one usually makes
rather detailed assumptions about the underlying organization structure on
which the individuals interact. The second line of approach is more in the
sense of cellular automata (CA) models with rather simple state variables and
local interaction dynamics. For example in the article by Wirl \cite{2} a
simple 1-dimensional deterministic cellular lattice automata model is used to
describe the propagation of corruption. Nevertheless, as is well known in
CA-modelling, the global dynamical picture can be highly complex and nontrivial.

Up to now all these attempts did not take into account the complex network of
social relationships as the underlying structure for the spread of corruption.
In this article we will present a model for the spread of corruption on
complex networks in the spirit of epidemiology. The model describes aspects of
the evolution of corruption in a virtual population and incorporates some
basic universal features of corruption. The local interaction dynamics of the
model is similar to cellular automata but "lives" not on a lattice type graph
like most of the CA-models but on complex networks.

Considering corruption as a nonstandard epidemic process relies on the
plausible assumption that corruption rarely emerges out of nothing but is
usually related to some already corrupt environment which may "infect"
susceptibles. Of course the spontaneous decision of somebody to act corruptly
is possible and could easy be handled in the model as an external weak source
of infection. One of the very special features in corruption propagation which
differs from what is used in describing classical epidemic processes is the
threshold like dependence of the local transition probabilities. By this we
mean that a noncorrupt individual gets infected with high probability if the
number of corrupt individuals in the group of his direct social contacts
(encoded as the set of neighbors in a "friendship" or acquaintance graph)
exceeds a certain threshold number. Otherwise if the number of corrupt
individuals in somebodies social neighborhood is below that threshold value
there is only a small probability to get corrupt via such "local"
interactions. The second main difference to classical epidemic processes is
the mean field dependence of the corruption process. By this we mean that an
individual can get corrupt just because there is a high prevalence (or
believed prevalence) in the society even when there is no corruption in the
local neighborhood. There is another interesting mean field term entering the
game, namely the society strikes back to corruption with an efficiency
proportional to the fraction of the noncorrupt people. Both mean field terms
are nonlinear and together with the local propagation mechanisms they give
rise to a rather complex dynamical picture.

\section{What is corruption?}

Corruption is a substructure of human social interaction. Common sense
associates corruption mainly with a deviation from fair play interaction in
the development of social relations. Clearly what is meant be fair play
depends on the cultural context of a given population/society. This vague
description of corruption is in the spirit of sociology and psychology and
differs from the more narrow corruption concepts usually considered in
economics or political sciences. There, corruption is mainly seen as a misuse
of public power to gain profit in a more or less illegal way. In any case,
corruption has many different faces in its concrete appearance and no single
model approach will be able to describe the whole picture in an adequate way.
But this does not at all imply that mathematical models are useless in this
situation. They can provide a substantial improvement in our understanding of
corruption as long as one clearly defines the aim and limitations of the taken approach.

For the model approach developed in this article we will use the notion of
corruption in the more general, first sense. More precisely our intention is
to describe changes in mind ranging from damming of corruption as a criminal
act to accepting corruption as an attractive option. Therefore in this paper
we do not introduce the group of state representatives or officials since we
assume that the essential changes in mind which allow corrupt acts happen long
before an individual is in the position to act corruptly. Empirical
investigations about motives and "typology" of corrupt actors (see \cite{33}
for results from case studies in Germany) have shown that the majority of
individuals involved in corruption affairs are highly educated, well
positioned with respect to social status and do not think to have done
something wrong, indicating the importance of mind changes prior to corrupt acts.
	
\begin{figure}
[ptb]
\begin{center}
\includegraphics[
height=2.6772in,
width=4.8799in
]%
{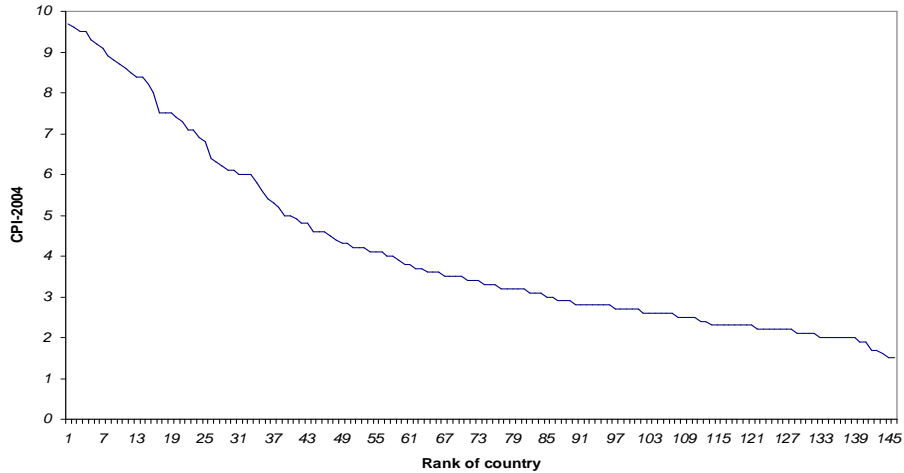}%
\caption{Corruption Perception index 2004 versus rank}%
\label{62}%
\end{center}
\end{figure}

There is a notorious problem in finding good empirical data which would allow
to estimate the prevalence of corruption. Probably the greatest effort over
the last years to measure the degree of corruption in various countries was
made by "Transparency International" (TA), a non profit group of individuals
and organizations which are highly concerned by the lack of sound data. Since
1995 they publish a yearly corruption report and a so called Corruption
Perception Index (CPI) \cite{34}. TA is well aware of the subjectivity in
peoples perception of corruption but hopes that the large number of cases
involved in the CPI averages out most of the bias. Figure \ref{62} gives a
CPI-rank plot of the 2004 date from TA. Note that a value of $10$ for the CPI
corresponds to the absence of corruption. For 2004 Finland holds the top
ranking and Germany is on place $15$ with an index of $8.2$.%

It is not our aim to explain the values of the CPI or other corruption data
sets, since this would require a semirealistic modelling of the social and
economical structure of individual countries which is completely illusionary
at the present stage of research. Rather we want to demonstrate which
scenarios are dynamically possible and whether there are phase transitions.

\section{Corruption as a generalized epidemic process}

In this section we first describe the basic setting for our model structure.
Refinements and more detailed aspects will be discussed later on. Due to the
common view, corruption is first of all a property of the relations between
individuals irrespectively which definition of corruption one uses. Since an act
of corruption requires that at least one of the participants in a corrupt
relation has a mental state which tolerates or even assigns a positive value
to (his personal view of) corruption we will focus mainly on the spread of
this mental state change (from not accepting to accepting corrupt acts as an
option for one's own activities). Therefore to discuss corruption as an
epidemic process in the afore mentioned sense it is useful to assign a
corruption property to the individuals themselves. In the simplest case we
just have a time dependent $0-1$ state variable $\omega\left(  x,t\right)  $
assigned to each individual, encoding whether the vertex is corrupt $\left(
1\right)  $ or not $\left(  0\right)  $ at time $t$ (of course more refined
scales for the degree of corruption are possible and will be discussed in a
forthcoming paper). The underlying structure on which corruption spreads is a
given finite graph $G$ from a random graph space $\mathcal{G}$ with fixed
vertex set $V=\{1,.....,n\}.$ Furthermore we consider in this article only
stationary graphs with no changes in time on the underlying graph structure
(the study of corruption on evolutionary graphs requires a paper in its own).
The dynamics is specified by conditional transition probabilities $\left(
p_{ij}\left(  x\right)  \right)  $ which depend mainly on the states on
$B_{1}\left(  x\right)  =\left\{  y:d\left(  x,y\right)  \leq1\right\}  $ and
a meanfield term reflecting the influence of the total prevalence of
corruption in the society. Here $d\left(  .,.\right)  $ is the usual graph
metric on $G\in\mathcal{G}$ and $d\left(  x\right)  $ is the degree of $x$. We
define $b_{t}:=\frac{1}{N}\sum\limits_{y\in V}\omega\left(  y,t\right)  $ as
the density of corruption at time $t$. The standing assumptions on $\left(
p_{ij}\left(  x\right)  \right)  $ are the following:
\begin{align}
p_{01}\left(  x\right)   &  =\Pr\left\{  \omega\left(  x,t+1\right)
=1\mid\omega\left(  x,t\right)  =0\right\}  =f_{x}\left(  \sum\limits_{y\sim
x}\omega\left(  y,t\right)  \right)  +\beta\left(  x\right)  \cdot b_{t}%
^{2}\nonumber\\
p_{10}\left(  x\right)   &  =\Pr\left\{  \omega\left(  x,t+1\right)
=0\mid\omega\left(  x,t\right)  =1\right\}  =\gamma\left(  x\right)
\cdot\left[  1-b_{t}\right]
\end{align}
in other words the probability to become corrupt depends only on the local
prevalence of corruption among the neighbors and the mean corruption in the
society and individuals who became corrupt can cure from corruption with a
rate proportional to the density of the noncorrupt individuals in the society.
In classical i.i.d. epidemics one would have the following as functional
dependence for the local part of the conditional probabilities: $f\left(
k\right)  =1-\left(  1-\varepsilon\right)  ^{k}$ which is for small
$\varepsilon$ and $k$ proportional to $\varepsilon k.$ For corruption the
function $f$ is more like in voter models, that is below a critical value
$\Delta\left(  x\right)  $ of the number of corrupt individuals in
$B_{1}\left(  x\right)  $ the value of $f$ is close to zero and above
$\Delta\left(  x\right)  $ it is a number $\alpha\left(  x\right)  $ much
larger then zero. Due to this property local clustering can force the
epidemics to spread whereas in classical epidemic processes high clustering
slows down the spread of an infection due to reinfection of the already
infected. We want to illustrate this by two simple examples.

\begin{example}
The simplest, almost trivial example is the $\mathbb{Z}^{1}$ lattice with
additional edges to the next-nearest neighbors. Setting $f\left(  1\right)
=\mu>0$ and $f\left(  i\right)  =1$ for $i>1$ it is easy to see that there is
a nonzero probability for infecting all vertices starting with one infected
individual at time $0.$
\end{example}

\begin{example}
The infection function $f$ will be the same as in example 1. We start with a
regular tree of degree 3. Replacing each vertex by a triangle and gluing the
triangles along the former edges of the regular tree gives a regular graph of
degree 4 where the triangle corners act now as the new vertices. In each
neighbor pair of triangles $\left(  A,B\right)  $ (that are the triangles
which have a common vertex) we form an edge randomly between the set of
vertices lying in $A\setminus B$ and $B\setminus A$ (see Fig. 1).%
\begin{figure}
[ptb]
\begin{center}
\includegraphics[
height=177.5pt,
width=273.6875pt
]%
{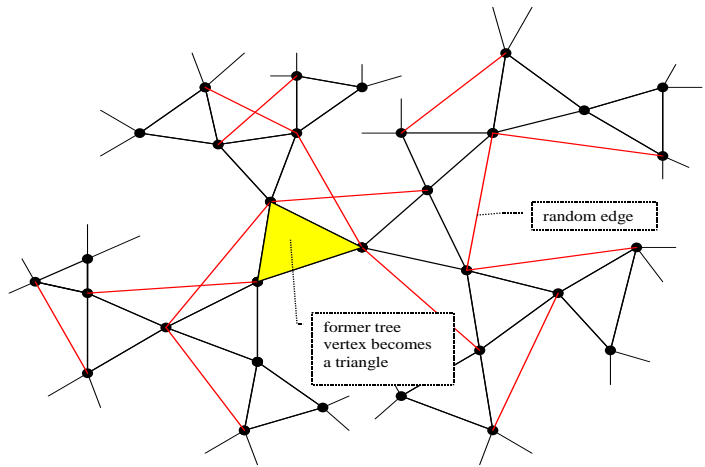}%
\caption{A highly clustered network with underlying tree structure. }%
\label{1}%
\end{center}
\end{figure}
Once a triangle is infected the corruption jumps to all the three neighbor
triangles due to the extra random edge present between each neighbor pairs of
triangles. Hence again we have a nonzero probability that the whole graph
becomes infected.
\end{example}

In the above examples we have used a very simple and somehow extreme form of
the infection function $f.$ In the following we will investigate the situation
for two canonical subclasses of infection functions. We say that $f$ is a
vertex independent, fixed threshold infection function if there is a $\Delta$
such that $f\left(  i\right)  =\varepsilon$ for $0<i<\Delta$ and $f\left(
i\right)  =\alpha\gg\varepsilon$ for $i\geq\Delta.$ For the second class of
functions we assume the threshold to be degree dependent. Namely we call
$f_{x}$ a vertex dependent, relative threshold function if for some $\delta
\in\left(  0,1\right)  $ we have $f_{x}\left(  i\right)  =\varepsilon$ for
$0<i<\delta d\left(  x\right)  $ and $f\left(  i,x\right)  =\alpha
\gg\varepsilon$ for $i\geq\delta d\left(  x\right)$. Furthermore we say that
$f$ is a voter-type infection function if $f\left(  i,x\right)  =\varepsilon$
for $i<\left\lfloor \frac{1}{2}d\left(  x\right)  \right\rfloor $ and
$f\left(  i,x\right)  =\alpha$ for $i\geq\left\lfloor \frac{1}{2}d\left(
x\right)  \right\rfloor$. In this paper we will mainly investigate the spread
of corruption for the fixed threshold case.

To distinguish between the different ways in which an individual can become
corrupt we will speak about the $\alpha,\beta,\varepsilon$ or $\gamma-$
process. For convenience of the reader we give in tabular 1 a summary of the
different processes.%

\begin{table}[tbp] \centering
\small{
\begin{tabular}
[c]{|l|l|l|}\hline
\emph{process name} & \ \ \ \ \ \ \ \ \ \ \ \ \ \ \ \ \emph{characteristic} &
\emph{typical value}\\\hline
$\alpha$ - process & $%
\begin{array}
[c]{c}%
\text{the local transmission process for}\\
\text{\# of corrupt neighbors}\geq\Delta
\end{array}
$ & $\alpha>>\varepsilon,\beta,\gamma$\\\hline
$\beta$ - process & $%
\begin{array}
[c]{c}%
\text{the mean field transmission process due to the}\\
\text{total prevalence or perception of corruption}%
\end{array}
$ & $\varepsilon<\beta<\gamma$\\\hline
$\gamma$ - process & $%
\begin{array}
[c]{c}%
\text{the corruption recover/elimination process}\\
\text{due to the fight of the society against corruption}%
\end{array}
$ & $\beta\leq\gamma<\alpha$\\\hline
$\varepsilon$ - process & $%
\begin{array}
[c]{c}%
\text{the classical local epidemic process for}\\
\text{\# of corrupt neighbors }<\Delta
\end{array}
$ & $\varepsilon<<\alpha,\beta,\gamma$\\\hline
\end{tabular}}%
\caption{the different processes for the corruption dynamics\label{key}}%
\end{table}%

Note that in contrast to standard voter models we do not have the possibility
of a locally induced backflip from the corrupt state to the noncorrupt. A kind
of quenched disorder could easily be introduced by randomizing the relevant
parameters individually but this will be the subject of a forthcoming paper.
Generalizations of classical epidemic dynamics to processes with a local
threshold have recently also been studied in the context of models of
contagion (see \cite{25} and references therein) but not yet been mixed with
global mean field processes.

\section{The structure of social networks}

In the last 10 years there has been an enormous progress in understanding the
fine structure of social networks. This is mostly due to the availability of
large data sets for some special social networks like E-mail correspondence,
coauthorship network in scientific publications and movie actor networks to
name just a few prominent examples. All these networks of social relations
share three remarkable properties of the associated graph which are: 1) the
diameter scales at most logarithmically in size 2) the graphs have a very
high, asymptotically non-vanishing clustering coefficient- in other words the
graphs are locally far from being tree-like 3) the degree distribution follows
a power law (scale-free graphs). Properties 2 and 3 have striking consequences
for the spread of corruption as will be discussed later on.

There exists meanwhile a large collection of algorithms to generate complex
networks with the above mentioned properties.

A widely used quantity to measure the local clustering is the triangle number
$A\left(  x\right)  :=\#\left\{  triangles\text{ }containing\text{ }x\right\}
$ and it's averaged value $\bar{A}$. A natural generalization is the $k-$
clique number $C_{k}\left(  x\right)  $ defined as the number of complete
graphs of order $k$ containing $x.$ In social network graphs $A\left(
x\right)  $ is usually proportional to $d\left(  x\right)  $ and $\bar{A}$
becomes independent of the population size for large $N$ and stays bounded
away from zero. Another very remarkable property of real networks is the
power-law distribution for the degree. By an asymptotic power-law distribution
for a discrete random variable $d$ we denote every functional behavior of the
form $\Pr\left\{  d=k\right\}  =k^{-\lambda+o_{k}\left(  1\right)  }$ with
exponent $\lambda>1.$ Most real networks have exponents between 2 and 4 (see
\cite{1} for an excellent overview). Classical epidemic processes on such
graphs have been studied by many authors, and perhaps the most astonishing
result in this context is the absence of an epidemic threshold in case the
exponent is below $3$ (\cite{3}). This phenomenon is related to the existence
of a massive center of size independent diameter induced by the high number of
hubs (vertices with an exceptional large degree). Hubs play also a significant
role for $\alpha$ - process as will be explained in the next section.

One of the main differences between corruption epidemics and classical
epidemics is the different effect of clustering on the epidemic threshold and
the total number of infected individuals. In the classical situation any
epidemics will be slowed down by the presence of local cycles due to the high
probability of reinfection. In corruption epidemics local clustering may speed
up the propagation of corruption due to the nonlinear dependence of the
infection probability on the number of infected neighbors as was already
demonstrated in example 1 in the previous section. In the next section we will
give two further examples where the strength of this effect can explicitly be computed.

\section{Phase transitions for the $\alpha$ - process}

In this section we want to look at some threshold properties associated to the
$\alpha-$ process. We are still far from a good understanding of the
quantitative picture of this kind of processes for a given type of graph,
which is mainly a consequence of our lack of knowledge how to handle graphs
with high local clustering in a mathematical satisfactory way. In this section
we want to state just some general observations and numerical results
concerning the spread of threshold-like dynamics. Furthermore we will analyze
two examples of tree-like graphs which might serve as an illustration. A more
careful mathematical analysis of $\alpha-$ processes requires a paper in its own.

\begin{figure}
[ptb]
\begin{center}
\includegraphics[
height=2.3321in,
width=3.0966in
]%
{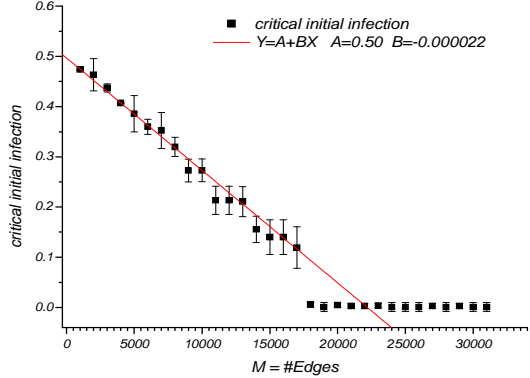}%
\caption{numerical estimation of the critical initial density $b_{0}^{c}$ as a
function of the number of edges $M$ for the following parameter values:
$\Delta=5;\alpha=0.35;\beta=0.08;\gamma=0.04;\varepsilon=0.005N=4000$
.Vertical segments are errorbars over 20 runs.}%
\label{52}%
\end{center}
\end{figure}

One of the remarkable differences between a classical epidemic process and a
process based on local threshold dynamics is the dependence on the initial
number of "infected" vertices in the latter case. Classical epidemics does not
know such things- either an epidemic process is overcritical (reproduction
number $R_{0}>1$) and a single initial infected vertex infects with with
positive probability a positive fraction of the whole population, or the
process is below criticality ($R_{0}<1$) and all infected will die out
respectively become healthy. In corruption epidemics both parts - the mean
field process as well the local $\alpha-$ process - can have phase transitions
with respect to the initial number of corrupt vertices. That means, there is
critical initial density of corrupt vertices $b_{0}^{c}$ such that for initial
densities below $b_{0}^{c}$ the number of infected stays as it is or goes down
to zero. Above $b_{0}^{c}$ the entire population becomes corrupt with high
probability. As an illustration we give in Fig. \ref{52} the dependence of
$b_{0}^{c}$ on the edge density $\frac{M}{N}$ on a classical random graph
space $\mathcal{G}\left(  N,M\right)  $ with $N$ vertices and $M$ edges.%

\begin{figure}
[ptb]
\begin{center}
\includegraphics[
height=2.3321in,
width=3.0966in
]%
{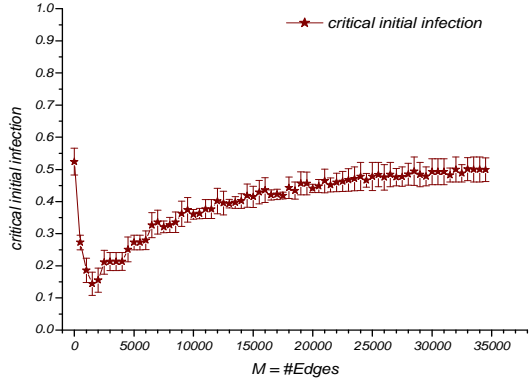}%
\caption{critical initial density $b_{0}^{c}$ as a function of the number of
edges $M$ for relative delta and parameter values: $\Delta\left(  x\right)
=\lceil0.8\cdot d\left(  x\right)  \rceil;\alpha=0.35;\beta=0.08;\gamma
=0.04;\varepsilon=0;N=4000$. Vertical segments are errorbars over 20 runs.}%
\label{55}%
\end{center}
\end{figure}

\begin{figure}
[ptbptb]
\begin{center}
\includegraphics[
height=1.8703in,
width=3.1935in
]%
{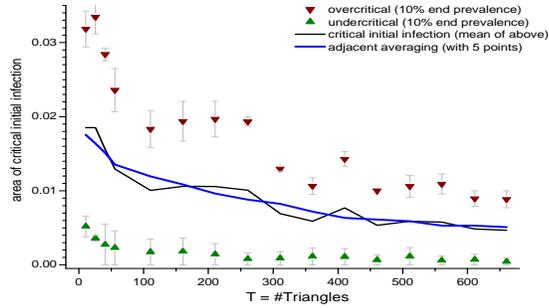}%
\caption{Critical density $b_{0}^{c}$ versus triangle density for the
parameter values: $\Delta=2;\alpha=0.3;\beta=0.08;\gamma=0.04;\varepsilon
=0.005;N=1000;M=2000.$ }%
\label{99}%
\end{center}
\end{figure}

Although in this paper we mainly concentrate on the case of absolute threshold
values $\Delta$ we give for comparison in Fig.\ref{55} the edge density
dependence result for a relative, degree dependent threshold $\Delta\left(
x\right)  =\lceil0.8\cdot d\left(  x\right)  \rceil$. There is still a
critical density but its value increases with the edge density since the mean
threshold increases now proportional to the mean degree.%
As already mentioned in section 3 one expects that the presence of
clustering (respectively many triangles) decreases the critical density
$b_{0}^{c}$ since the $\alpha$- process can propagate more easily. In Fig.
\ref{99} the effect of the increase of the triangle number is clearly to see.
Here we used a modified $\mathcal{G}\left(  N,M\right)  $ random graph space
where randomly triangles are added (keeping the total number of edges
constant). The threshold value $\Delta$ was chosen to be $2$ since for higher
$\Delta$ one has to add higher order complete subgraphs instead of triangles.%

The next figure (Fig. \ref{300}) shows the dependence of the critical density
on $\Delta.$ The two curves represent the threshold values for an
end-prevalence of $10$ respectively $90$ percent. Since the mean degree in
this simulation is about $6.5$ one has a vanishing contribution of the
$\alpha$- process above $\Delta=8$. The critical threshold $b_{0}^{c}$ stays
than essentially at a value given by the mean field process (see next section
for details).
\begin{figure}
[ptb]
\begin{center}
\includegraphics[
height=2.3321in,
width=3.1537in
]
{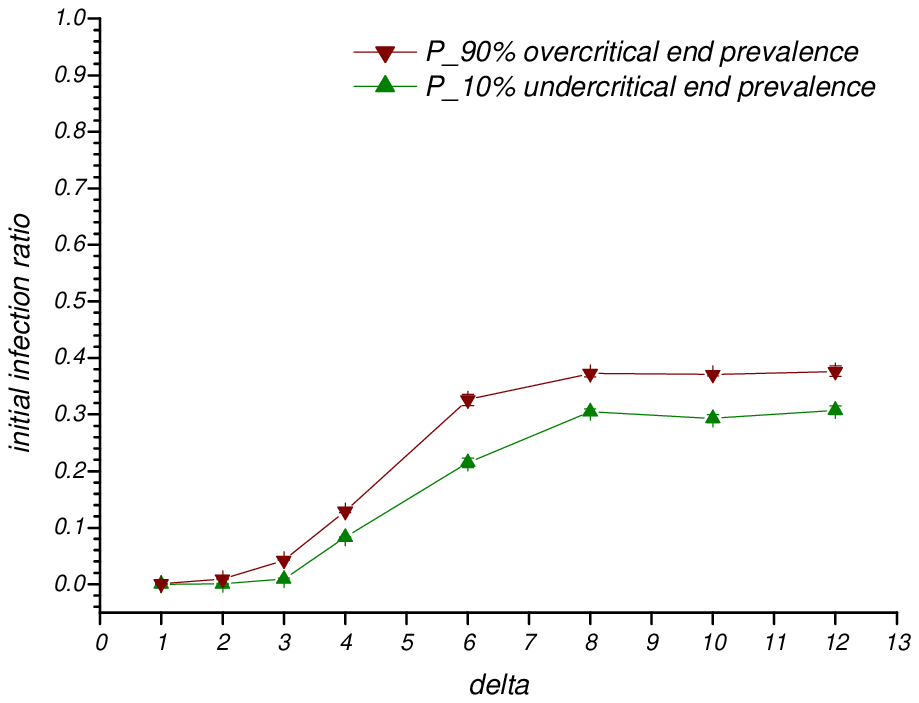}
\caption{Lower and upper bounds for the critical density $b_{0}^{c}$ as a
function of $\Delta$ for the following parameter values: $N=1500;M=5000;\alpha
=0.35;\beta=0.08;\gamma=0.04;\varepsilon=0.005$}%
\label{300}%
\end{center}
\end{figure}
To get an impression of the contribution of the different kind of processes
(local $\alpha$ and $\varepsilon$, global $\beta$ and $\gamma$ - for details
see next section) to the end-prevalence we give in Fig. \ref{400} the
accumulated number of state changes caused by each of the subprocesses till
saturation. For small values of $\Delta$ the $\alpha$- process dominates all
others.%
\begin{figure}
[ptbptb]
\begin{center}
\includegraphics[
height=2.3321in,
width=3.5169in
]%
{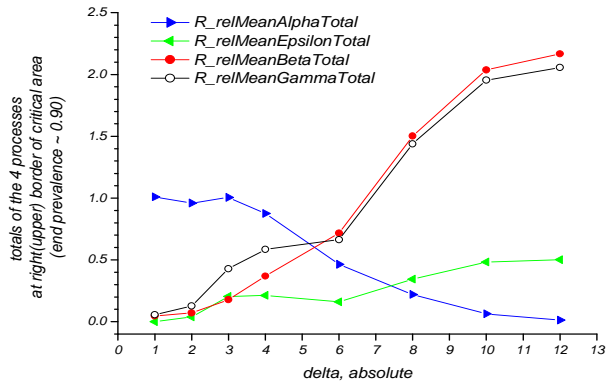}%
\caption{Total number of state changes splitted according to the different
subprocesses as a function of $\Delta$ for the same parameter values as in
Fig.\ref{300}.}%
\label{400}%
\end{center}
\end{figure}

\begin{figure}
[ptb]
\begin{center}
\includegraphics[
height=2.3321in,
width=3.1805in
]%
{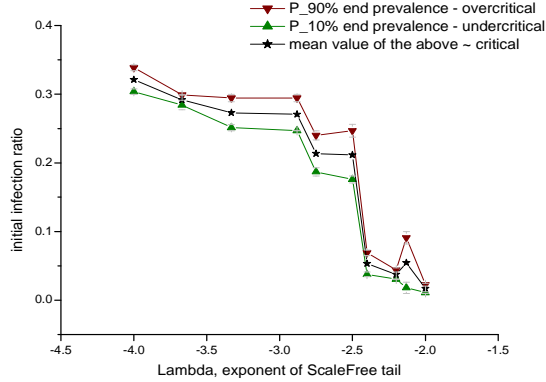}%
\caption{$b_{0}^{c}$ as a function of the exponet $\lambda$ in a scale-free
degree distribution with parameters: $N=20000;M=50000;\Delta=5;\alpha
=0.35;\beta=0.08;\gamma=0.04;\varepsilon=0$}%
\label{56}%
\end{center}
\end{figure}

\begin{figure}
[ptbptb]
\begin{center}
\includegraphics[
height=2.2941in,
width=3.2903in
]%
{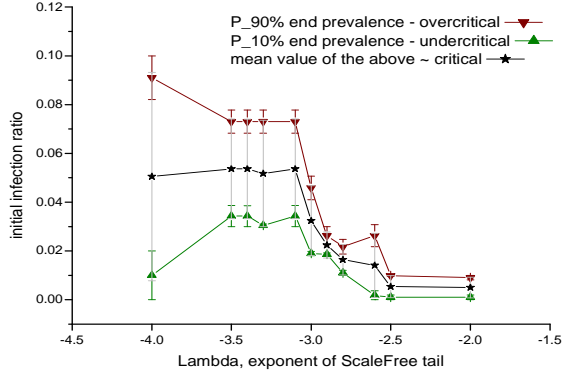}%
\caption{upper and lower bounds for $b_{0}^{c}$ as a function of the exponet
$\lambda$ in a scale-free degree distribution with parameters:
$N=20000;M=30000;\Delta=2;\alpha=0.35;\beta=0.08;\gamma=0.04;\varepsilon=0$}%
\label{57}%
\end{center}
\end{figure}

\begin{figure}
[ptbptbptb]
\begin{center}
\includegraphics[
height=2.3148in,
width=3.2333in
]%
{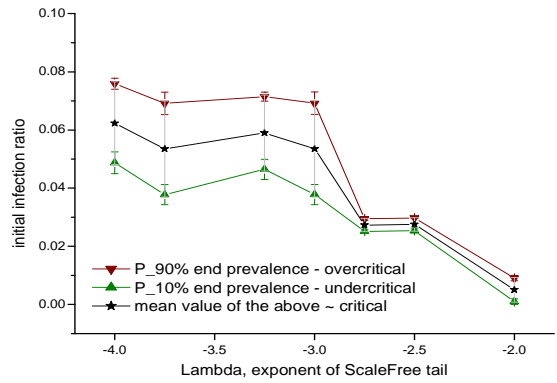}%
\caption{upper and lower bounds for $b_{0}^{c}$ as a function of the exponet
$\lambda$ in a graph with scale-free degree distribution, additive degree
correlation and parameters: $N=20000;M=30000;\Delta=2;\alpha=0.35;\beta
=0.08;\gamma=0.04;\varepsilon=0$}%
\label{58}%
\end{center}
\end{figure}

\begin{figure}
[ptbptbptb]
\begin{center}
\includegraphics[
height=2.3148in,
width=3.2333in
]%
{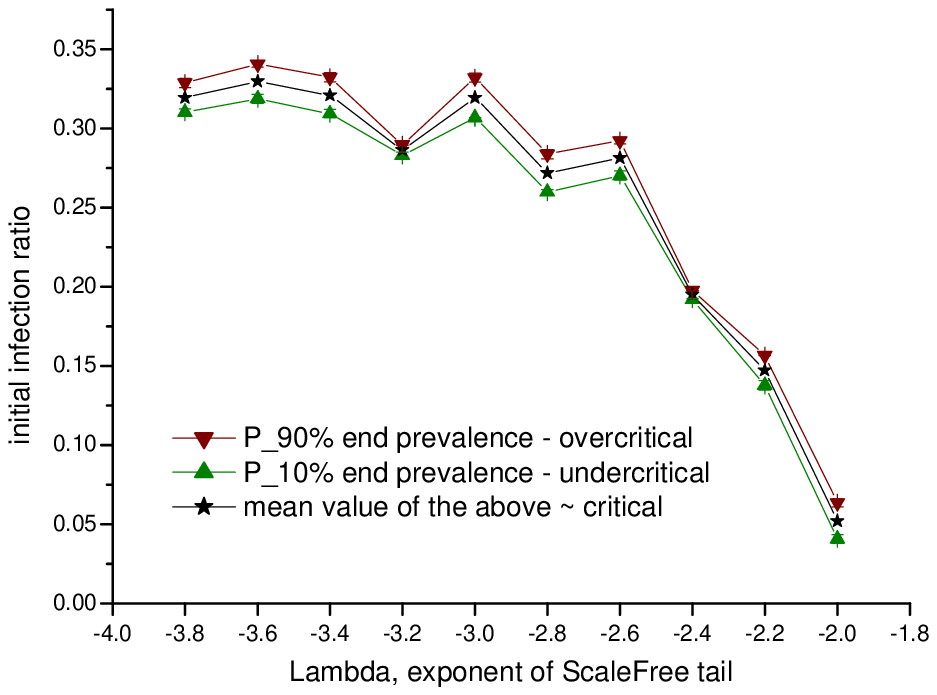}%
\caption{$b_{0}^{c}$ as a function of the exponet $\lambda$ in a scale-free
degree distribution with parameters: $N=20000;M=50000;\Delta=5;\alpha
=0.35;\beta=0.08;\gamma=0.04;\varepsilon=0$}
\label{999}%
\end{center}
\end{figure}

We turn now to a more theoretical consideration, namely which type of vertices
(type in the sense of degree and local clustering) are especially well suited
for the propagation of corruption via the $\alpha-$process. Assume we have
given a random scale free graph space $\mathcal{G}$ with $N$ vertices. We
further assume that there are two types of edges (according to the way they
where generated) the independent ones, generated at random with just
preferences to the degree (like the preferential attachment rule by
Albert\&Barabasi or the "Cameo-Principle" in \cite{7}) and local ones which
are relevant for the creation of triangles. Let further the (asymptotic)
degree distribution given by $\varphi\left(  k\right)  :=\Pr\left\{  d\left(
x\right)  =k\right\}  =\frac{B\left(  k_{0}\right)  }{k^{\lambda}}$ with
$k_{\min}\geq k_{0}$ and $B\left(  k_{0}\right)  $ the normalization constant.
The independent edges are generated with probability $p_{0}$ and each
individual generates $k_{0}>2$ edges by himself. From \cite{7} one knows that
the triangle number $A\left(  x\right)  $ is proportional to $d\left(
x\right)  $ with a proportionality constant $C\left(  p_{0}\right)$. A basic
quantity in highly clustered networks is the probability $q\left(  x\right)  $
that two random chosen elements from $N_{1}\left(  x\mid d\left(  x\right)
>1\right)  $ have a common edge. Since $A\left(  x\right)  \sim c_{T}d\left(
x\right)  $ one obtains for the conditional probability $q_{k}\left(
x\right)  :=\Pr\left\{  z\sim y\mid z,y\in N_{1}\left(  x\right)  \wedge
d\left(  x\right)  =k\right\}  \sim\frac{2c_{T}}{k-1}\sim\frac{2c_{T}}{k}$.
Assuming that the generation of triangles is a sufficiently independent
process one obtains for the conditional $l-$ clique number $\mathbb{E}\left(
C_{l}\left(  x\right)  \mid d\left(  x\right)  =k\right)  =\binom{k}%
{l-1}\left(  \frac{2c_{T}}{k}\right)  ^{\frac{\left(  l-1\right)  \left(
l-2\right)  }{2}}\sim\frac{k^{\left(  l-1\right)  \left(  1-\frac{l-2}%
{2}\right)  }}{\left(  l-1\right)  !}\cdot const.$ Here $C_{l}\left(
x\right)  $ is the number of $l$- cliques (complete graphs of order $l$)
containing $x$. For $l>4$ the power in the $k-$dependence gets negative and
hence the high degree vertices contribute almost nothing to the
Clique-clustering. Of course all this consideration rely on the assumption of
some kind of independence in the triangle-formation process. In any case this
results indicate that highly clustered medium degree vertices are especially
well suited for the spread of corruption. A similar kind of analysis can be
carried out for random graph models which have an intrinsic high probability
to generate local cliques e.g. intersection graphs (for an introduction to
random intersection graphs and comparison with Erd\"{o}s-Renyi random graphs
see \cite{23} and \cite{24}). The above arguments seem to support the
conjecture that in corruption epidemics the vertices from the tail of the
degree distribution play a less dominant role. This is indeed true in the case
of a relative, degree-dependent threshold where hubs are much more difficult to
infect than medium or low degree vertices. For absolute thresholds in the
$\alpha$- process the situation is more complex since for scale free degree
distributions with small exponents ($\lambda<3$) there are other mechanisms
than local clustering which can cause a radical dropdown of the critical
initial density. In Fig.\ref{56} and Fig.\ref{57} we give numerical results
for the relation between the critical density $b_{0}^{c}$ and the exponent
$\lambda$ keeping the edge density fixed. There is a clear phase transition
around $\lambda\sim2.3$ for $\Delta=5$ and $\lambda\sim2.9$ for $\Delta=2$.%
The explanation of this observation is closely related to a structural phase
transition in scale free random graphs at $\lambda=3$ - namely that for most
vertices $x$ an asymptotically positive fraction of all vertices has bounded
distance to $x$. To link this property with the $\alpha$ - process one has to
look more closely on the degree-degree correlation in scale-free graphs.
Depending on the choice of the model one can have very different correlations
like:%
\begin{align}
\Pr\left\{  x\sim y\mid d\left(  x\right)  =k\wedge d\left(  y\right)
=k^{\prime}\right\}   &  \simeq const\cdot\frac{k+k^{\prime}}{N}\text{
or}\label{111}\\
\Pr\left\{  x\sim y\mid d\left(  x\right)  =k\wedge d\left(  y\right)
=k^{\prime}\right\}   &  \simeq const\cdot\frac{k\cdot k^{\prime}}%
{N}\label{113}%
\end{align}
. Formula \ref{111} holds for instance for the Cameo - model (\cite{7})
whereas formula \ref{113} is valid for scale-free graphs generated via the
Molloy\&Reed algorithm (the later one represents the random graph space
containing all graphs with a given scale-free degree distribution equipped
with the uniform measure and was used for the simulations in Fig.\ref{56} and
\ref{57}). Evolutionary graphs like the Albert\&Barabasi model have usually
asymmetric and more complicated correlations. Since a detailed analysis of the
$\alpha$ - process for scale-free graphs is beyond the scope of this paper we
just give a heuristic outline why in graphs with a correlation as in formula
\ref{113} the threshold density $b_{0}^{c}$ tends to zero as $N\rightarrow
\infty$ for exponents $\lambda<3$. For fixed $b_{0}>N^{\frac{1}{\lambda}-\nu}$
and $\nu>0$ (note that the typical maximal degree is about $N^{\frac
{1}{\lambda}})$ it is obvious that vertices $x$ with $d\left(  x\right)  \geq
k_{0}>>\frac{\Delta}{b_{0}}$ get almost surely infected (as $N\rightarrow
\infty$) via the $\alpha$ - process as soon as $\gamma<\alpha$. Let $A_{k_{0}%
}$ be the set of such vertices. One the other side it follows from \ref{113}
that a vertex $y$ with $d\left(  y\right)  =k<k_{0}$ is linked to the set
$A_{k_{0}}$ with probability%
\begin{align}
q_{k} &  \sim1-\prod\limits_{k^{\prime}\geq k_{0}}^{k_{\max}\sim N^{\frac
{1}{\lambda}}}\left(  1-\frac{const\cdot k\cdot k^{\prime}}{N}\right)
^{const\cdot\frac{N}{\left(  k^{\prime}\right)  ^{\lambda}}}\\
&  \sim1-e^{-const\cdot\frac{k}{N}\sum\limits_{k^{\prime}\geq k_{0}}%
N\cdot\frac{k^{\prime}}{\left(  k^{\prime}\right)  ^{\lambda}}}\sim
1-e^{-const\cdot k\frac{1}{k_{0}^{\lambda-2}}}%
\end{align}
. Since $q_{k}$ is close to $1$ for $k>k_{0}^{\lambda-2}$ one has an almost
sure multiple linkage of vertices $y$ with $d\left(  y\right)  >k_{0}%
^{\lambda-2}<k_{0}$ to the set $A_{k_{0}}$. These vertices get now again
infected via the $\alpha$ - process. By iterating this procedure one may
arrive at an positive $N$- independent infection density $b_{t}>>b_{0}$ such
that the $\beta$ - process is overcritical and finally the whole vertex set
becomes corrupt. The mechanism described requires $N$ to be large and
therefore we conjecture that the difference to the numerical results depicted
in Fig.\ref{56} (phase transition at $\lambda<2.3$ instead of $3$) is due to
finite size effects. In the case of $\Delta=2$ (Fig.\ref{57}) the finite size
effects are smaller and the phase transition is closer to $3$. A similar
kind of arguments shows, that the expected path-length is finite for
$\lambda<3$. Namely since the expected number $S_{l}$ of vertices at distance
$l$ from a vertex $x$ with degree $k_{0}$ is approximately given by $\left(
const\right)  ^{l}\cdot\sum\limits_{k_{1},...,k_{l}}^{N^{\frac{1}{\lambda}}%
}\frac{k_{o}\cdot k_{1}\cdot k_{1}\cdot...\cdot k_{l-1}\cdot k_{l-1}\cdot
k_{l}}{\left(  k_{1}\right)  ^{\lambda}\cdot...\cdot\left(  k_{l-1}\right)
^{\lambda}}\sim$ $const\cdot k_{0}\cdot N^{\frac{\left(  l-1\right)  \left(
3-\lambda\right)  }{\gamma}}$ for $\lambda<3$ (note that this expression is
only valid for $l$ s.t.$\frac{\left(  l-1\right)  \left(  3-\lambda\right)
}{\gamma}\cdot\log k_{0}<1$). The essential diameter $diam_{e}$ (a large
fraction if the whole vertex set is within a ball of diameter $diam_{e}$) is
then given by the smallest $l$ such that $\frac{\left(  l-1\right)  \left(
3-\lambda\right)  }{\gamma}>1$ (for a more extensive discussion of the notion
of essential diameter see \cite{28}). For $\lambda=2$ one obtains therefore
$diam_{e}=3$. For $\lambda>3$ the essential diameter is no longer bounded but
growths logarithmically in $N$. It is interesting that the jump in the
critical density at $2.3$ in Fig.\ref{56} coincides with a jump in diameter
from $4$ to $5$. A small essential diameter can have fatal consequences for
corruption epidemics since most vertices are closely linked to hubs and, as
was outlined above, hubs are with high probability corrupt. A precise
estimation of the dependence of $b_{0}^{c}$ from $N,M$ and $\lambda$ requires
a careful discussion of the involved constants. For scale-free graphs with
additive degree correlation like Cameo-graphs one still has a bounded
essential diameter for exponents less than $3$. But the first argument about
chains of almost sure linkages from high degree to low degree vertex sets can
not be adopted. One expects therefore a higher value of the critical density
$b_{0}^{c}$. In Fig.\ref{58} and Fig. \ref{999} we give numerical results for a scale-free graph
with additive degree-correlation generated via a modified Molloy\&Reed
algorithm \footnote{In the usual Molloy\&Reed algorithm one generates $d\left(
x\right)  $ virtual vertices for each vertex $x$ and makes than a random
matching between the virtual vertices. Two vertices $x$ and $y$ are connected
by an edge if there is an edge between two corresponding virtual vertices. To
generate an additive degree correlation we mark $M$ virtual vertices as red
such that each vertex $x$ has at least one red and at most $C$ $\left\lceil\frac{M}{N}
\right\rceil $ red associated virtual vertices (if there are note too many vertices with very small degree, the constant $C$ can be choosen as $\left\lceil\frac{M}{N}
\right\rceil $). Than the marked red
virtual vertices are randomly matched with the unmarked ones.}%

To compare with the multiplicative case we have chosen the same parameters and
degree-distribution as for Fig.\ref{57}. There is a clear increase of
$b_{0}^{c}$ to observe but, although unlikely, it remains open wether there is
a vanishing threshold in the limit $N\rightarrow\infty$. For intermediate
couplings we still expect $b_{0}^{c}\left(  N\right)  \rightarrow0$ as $N$
diverges for $\lambda<\lambda_{c}\in\left(  2,3\right)  $ where $\lambda_{c}$
depends on thew concrete model. It is remarkable that low $\lambda$ and a
tendency to multiplicative correlation is mainly expected to hold in societies
with strong hierarchical structures of social dependencies e.g. dictatorships
(see \cite{13} for details), whereas democracies are characterized by less
strong degree correlation.

\begin{figure}
[h]
\begin{center}
\includegraphics[
trim=0.000000in 0.000000in -0.001260in 0.000000in,
height=1.8158in,
width=3.6172in
]%
{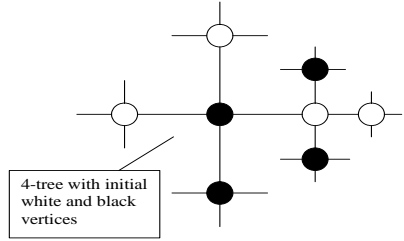}%
\caption{Segment of a regular infinite tree of order 4}%
\label{12}%
\end{center}
\end{figure}

Finally we will discuss two examples of graph structures where the critical
infection density can be explicitly be computed. The first one is a regular
infinite tree of degree 4 where of course no triangles are present (see Fig.
\ref{12}).%

\begin{figure}
[hptb]
\begin{center}
\includegraphics[
height=2.0684in,
width=3.4088in
]%
{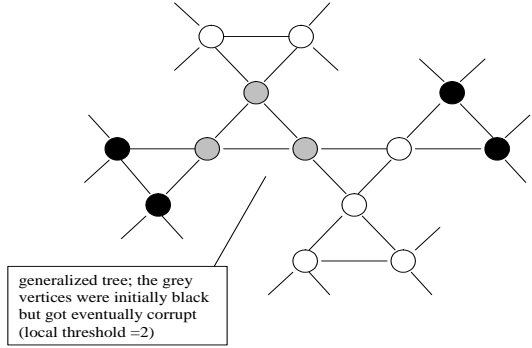}%
\caption{Segment of an infinite generalized tree (degree 4 and branching
number 3)}%
\label{13}%
\end{center}
\end{figure}

. The second structure is a regular infinite graph of again of degree 4 with
positive local cluster coefficient ($A\left(  x\right)  =2$) and a global
tree-like structure (see Fig.\ref{13}).%

\begin{figure}
[ptb]
\begin{center}
\includegraphics[
height=2.2500in,
width=3.1386in
]%
{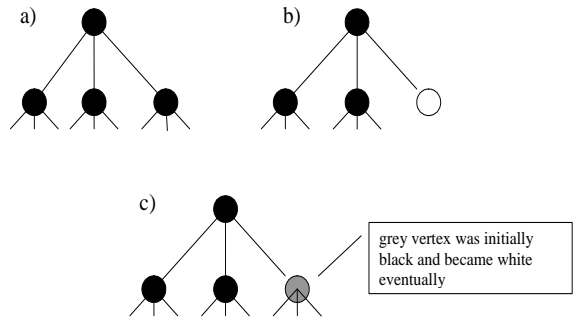}%
\caption{Different configurations in the neighborhood of the root vertex..
Black denotes vertices in an immune cluster and grey an initial black vertex
which became white. }%
\label{32}%
\end{center}
\end{figure}

In both cases an exact computation of the critical infection density is
possible. We give a short outline for the case of threshold value $\Delta=2$
and $\alpha=1$ (the case $\alpha<1$ requires more lengthy computations but can
be done in a similar fashion) and start with the case of the regular 4 tree.
An initial configuration is given by marking each vertex with probability $p$
as noncorrupt (black) and with probability $1-p$ as corrupt (white). We ask
for the critical probability $p_{c}$ such that for $p<p_{c}$ almost surely the
entire tree becomes white (corrupt) and for $p>p_{c}$ there remains an
infinite cluster of noncorrupt (black) vertices with probability one. Note
that no finite cluster of black vertices -that is a finite black subgraph
surrounded by white vertices- can survive so there are either infinite black
clusters or none. We call an invariant infinite black cluster immune. Since
$\Delta=2$ any vertex in an immune cluster must have at least three black
neighbors from that cluster. Denote by $T_{R}\left(  3\right)  $ the rooted
tree with outdegree $3$ (fixing a root gives a canonical direction to the
edges of the tree so it makes sense to speak about the outdegree of a vertex).
Every vertex has degree $4$ except the root which has degree $3$. Let $x$ be
the $p-$ dependent probability that the root is contained in an immune cluster
(as a subgraph of $T_{R}\left(  3\right)  $) conditioned that the root vertex
is initially black. By arguments from the general theory of branching
processes $x$ equals the largest solution of the following recursion equation
\begin{equation}
x=\underset{a)}{\underbrace{p^{3}x^{3}}}+\underset{b)}{\underbrace{3p^{3}%
x^{2}\left(  1-x\right)  }}+\underset{c)}{\underbrace{3p^{2}\left(
1-p\right)  x^{2}}}\ .
\end{equation}
Figure \ref{32} displays the different situations which enter the above
equation. The solutions are $\frac{1}{2p^{3}}\left(  \frac{3}{2}p^{2}\pm
\frac{1}{2}\sqrt{-8p^{3}+9p^{4}}\right)  $ and $0$. Since $-8p^{3}+9p^{4}%
\geq0$ is needed to have a positive nonzero solution we get for the critical
probability $p_{c}=\frac{8}{9}\simeq\allowbreak0.888\,89.$%

\begin{figure}
[ptb]
\begin{center}
\includegraphics[
height=3.2752in,
width=2.9627in
]%
{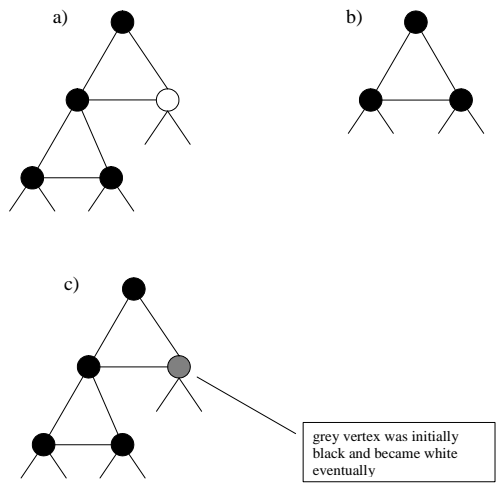}%
\caption{The local picture around the root vertex in $T_{R}\left(  2,1\right)
$}%
\label{42}%
\end{center}
\end{figure}

In a similar fashion one can derive a recursion equation for the generalized
tree case. For that let $T_{R}\left(  2,1\right)  $ be the rooted generalized
tree shown in Fig.\ref{42}.
To every vertex is attached an outgoing triangle, hence the degree of a vertex
is $4$ except the root which has degree $2$. To settle the question about
$p_{c}$ for the original generalized tree it is enough to analyze the
corresponding problem for $T_{R}\left(  2,1\right)  $. Again let $x$ be the
probability that the root vertex is in an immune cluster conditioned that the
root is initially black. One gets the following recursion equation%
\begin{equation}
x=\underset{b)}{\underbrace{p^{2}x^{2}}}+\underset{c)}{\underbrace{2p^{4}%
x^{2}\left(  1-x\right)  }}+\underset{a)}{\underbrace{2p\left(  1-p\right)
p^{2}x^{2}}}%
\end{equation}
(see Fig.\ref{42}). The solutions are $\frac{1}{2p^{4}}\left(  \frac{1}%
{2}p^{2}+p^{3}\pm\frac{1}{2}\sqrt{-7p^{4}+4p^{5}+4p^{6}}\right)  $ and $0$.
Again since $-7p^{4}+4p^{5}+4p^{6}\geq0$ is needed to get a positive nonzero
solution we get for the critical probability $p_{c}=\sqrt{2}-\frac{1}{2}%
\simeq\allowbreak0.914\,21.$ That means the presence of clustering in this
example lowers the critical initial density needed to infect the whole graph
by almost a factor of $\frac{3}{4}.$

The study of the regular $4-$tree generalizes easily to the case of regular
$n+1-$ trees ($n>2$). The recursion equation in this case is
\begin{equation}
x=p^{n}x^{n}+np^{n}x^{n-1}\left(  1-x\right)  +np^{n-1}\left(  1-p\right)
x^{n-1}\ .
\end{equation}
A straightforward but lengthy computation gives for the critical probability%
\begin{equation}
p_{c}=\frac{\left(  n-1\right)  ^{2n-3}}{n^{n-1}\left(  n-2\right)  ^{n-2}%
};n>2\ .
\end{equation}
In the special case of a $3-$ tree ($n=2$) one obtains $p_{c}=\frac{1}{2}$.
For completeness we give without proof the formula for the computation of the
critical probability in case of a rooted random tree with arbitrary outdegree
distribution. Let $g\left(  z\right)  =\sum\limits_{i\geq2}a_{i}z^{i}$ be the
generating function for the outdegree; that is $a_{i}$ is the probability that
a random chosen vertex has outdegree $i$ (and hence total degree $i+1$). The
critical probability $p_{c}$ is given by the smallest $p$ such that the
equation
\begin{equation}
\frac{z}{p}=\left(  1-z\right)  g^{\prime}\left(  z\right)  +g\left(
z\right)
\end{equation}
has a positive real solution.

A careful reader may have noticed that there is a big structural difference
between the generalized tree in example 2 of section 3 and the generalized
tree just discussed above. Namely the graph of the first example has the
property, that any two vertices can be linked by a chain of triangle where
neighbor triangles always have a common edge. The graphs in the examples of
this section do not have this property since neighboring triangles have only a
common vertex. For threshold values $\Delta>2$ one has to consider chains of
$\Delta+1$ cliques. We say that a graph is well $k$ -linked if any pair of
vertices can be linked by a chain of complete graphs of order $k$ such that
all neighboring $k-$ cliques have a $k-1$- clique in common. For well $k$-
linked graphs the critical density $b_{c}^{0}$ is zero (a finite number of
initially infected vertices can already infected a positive fraction of the
vertex set) for $\alpha-$processes with $\Delta<k$ whereas for graphs which
are not well linked one needs a positive critical density.

The above study on trees or generalized trees is insofar important as in most
random graph models used for complex networks one has as a tree or generalized
tree as the typical local structure around a random chosen vertex. Furthermore
the dependence of the corruption dynamics on graph properties like edge
density or degree distribution is in large parts of the parameter space
entirely caused by the $\alpha-$ process.

\section{ State and individual (mean field $\beta-$ and $\gamma-$ process)}

In this section we want to have a closer look at the mean field dependence of
the corruption process. To gain some insight in the possible type of behavior
we start with some simple assumptions which will be refined later on. Again we
will argue in a discrete time model but the transition to continuos time makes
no problem and gives the same results. Let $b_{t}$ the density of corrupt
people at time $t.$ We assume that the affinity for an individual to change
its behavior from noncorrupt to corrupt increases proportional to the
corruption prevalence. Furthermore to become really corruptly minded an
individual has to overcome some fear which we put proportional to $\left(
1-b_{t}\right)$. Formally this reads as $\Pr\left\{  \omega\left(
x,t+1\right)  =1\mid\omega\left(  x,t\right)  =0\right\}  =\beta b_{t}^{2}$
with $\beta\in\left[  0,1\right]$. Corrupt individuals can recover due to
state and police effects (uncovering, fear etc.). Again it seems reasonable to
assume that the probability to recover is proportional to $1-b_{t}$ since only
the noncorrupt part of a society is willing to fight corruption. Formally we
will assume that $\Pr\left\{  \omega\left(  x,t+1\right)  =0\mid\omega\left(
x,t\right)  =1\right\}  =\gamma\left(  1-b_{t}\right)  $ with $\gamma
\in\left[  0,1\right]$. This gives
\begin{align}
b_{t+1}  &  =\left(  1-b_{t}\right)  \beta b_{t}^{2}+b_{t}-b_{t}\gamma\left(
1-b_{t}\right) \nonumber\\
&  =b_{t}\left(  1-\gamma\right)  +b_{t}^{2}\left(  \gamma+\beta\right)
-b_{t}^{3}\beta\label{100}%
\end{align}
with the two obvious fixed points $0$ and $1.$ For $\beta$ $\neq0$ there is a
third intermediate fixed point $b^{\ast}:=\frac{\gamma}{\beta}.$ An
interesting phenomena happens for parameter pairs $\left(  \beta
,\gamma\right)  $ s.t. $\gamma<\beta$ since under this conditions both fixed
points at $0$ and $1$ are locally stable. Hence there are two basins of
attraction- one for $0$ and one for $1-$ with $b^{\ast}$ as the boundary
point. In other words, if the initial percentage of corruption is less
$b^{\ast}$ corruption stays under control whereas for an initial value larger
$b^{\ast}$ things run out of control and a corruption collapse takes place. Of
course this mean field part of the model is still very simplistic and one
should not expect any quantitative fit with empirical data. But the
qualitative statement seems to be quite stable with respect to modifications.
For instance there are good reasons to believe that neither the mean field
infection nor the mean field recover process are linear in $b_{t}.$

We want to end this section with a small modification of the mean field
"Ansatz" where we include social weights. This is a natural and meanwhile very
common approach in network dynamics and can easily be adopted to the
corruption model. In the above argumentation on the attraction of becoming
corrupt it is plausible to assume that corrupt individuals with high social
influence have a stronger influence on the mean field probability to get
corrupt than individuals with low social importance. A similar argument holds
for the recover probability. As a simple measure for social strength we use
the degree of the vertices since high degree vertices are more likely to play
a dominant social role than low degree vertices. Formally we introduce the
weighted density $b_{t}^{w}$at time $t$ as
\begin{equation}
b_{t}^{w}:=\frac{\sum\limits_{k}I_{t}^{\left(  k\right)  }d_{k}}%
{\sum\limits_{k}\left(  d_{k}\right)  ^{2}}%
\end{equation}
where $d_{k}$ is the number of vertices with degree $k$ and $I_{t}^{\left(
k\right)  }$ the number of corrupt (state $1$) vertices with degree $k$ at
time $t$. The mean field equation for group $k$ is now given by%
\begin{equation}
I_{t+1}^{\left(  k\right)  }=\beta\left(  b_{t}^{w}\right)  ^{2}\left(
d_{k}-I_{t}^{\left(  k\right)  }\right)  +I_{t}^{\left(  k\right)  }%
-\gamma\left(  1-b_{t}^{w}\right)  I_{t}^{\left(  k\right)  }%
\end{equation}
. Multiplying the last equation by $\frac{d_{k}}{\sum\limits_{k}\left(
d_{k}\right)  ^{2}}$ and summing over $k$ gives
\begin{equation}
b_{t+1}^{w}=\left(  1-b_{t}^{w}\right)  \beta\left(  b_{t}^{w}\right)
^{2}+b_{t}^{w}-b_{t}^{w}\gamma\left(  1-b_{t}^{w}\right)
\end{equation}
which is the same as equation (\ref{100}). Therefore the introduction of
social weights does not add anything new to the dynamical picture. There is of
course a difference in the interpretation since a small real initial
prevalence of corruption can give rise to a high initial value of $b_{0}^{w}$
as soon as the corruption is concentrated at the high degree vertices. Here
also a difference between scale free networks and classical random networks is
seen since in the scale free case high degree vertices (hubs) are much more
frequent than in the classical case.

\section{Interaction between the mean field process and the local threshold
dynamics ($\beta+\gamma$ versus $\alpha$)}

In this paragraph we will investigate some aspects of the interplay between
the mean field process described in the previous section and the local,
threshold dependent, corruption propagation. For $\alpha>\gamma$ there is a
core infected component generated via the $\alpha-$ process. To gain some
insight how such a core infected part of the population changes the mean field
dynamics we will assume that a certain fraction, say $a,$ of the population is
permanently infected and resistant to the $\gamma-$deletion process. Denoting
by $q_{t}=b_{t}-a$ the density in the noncore part of the population (the
normalization here is still with respect to the total population size) we get
the following mean field dynamics:%
\begin{align}
q_{t+1}  &  =\left(  1-a-q_{t}\right)  \beta\left(  q_{t}+a\right)  ^{2}%
+q_{t}-q_{t}\gamma\left(  1-a-q_{t}\right) \nonumber\\
&  =a^{2}\beta-a^{3}\beta+q_{t}\left(  2a\beta-\gamma+a\gamma-3a^{2}%
\beta+1\right)  +\nonumber\\
&  +q_{t}^{2}\left(  \beta+\gamma-3a\beta\right)  -\beta q_{t}^{3}%
\end{align}
. Since the state where all individuals are infected is stationary we get the
following set of fixed points:%
\[
\left\{  -a+1,\frac{1}{\beta}\left(  \frac{1}{2}\gamma-a\beta-\frac{1}{2}%
\sqrt{-4a\beta\gamma+\gamma^{2}}\right), \frac{1}{\beta}\left(  \frac{1}%
{2}\gamma-a\beta+\frac{1}{2}\sqrt{-4a\beta\gamma+\gamma^{2}}\right)  \right\}
\]
. For $-4a\beta\gamma+\gamma^{2}<0$ there are no real fixed points except
$q^{\ast}=-a+1$ which becomes globally stable under this condition. Since we
have a polynomial of degree 3 we get $\frac{1}{\beta}\left(  \frac{1}{2}%
\gamma-a\beta+\frac{1}{2}\sqrt{-4a\beta\gamma+\gamma^{2}}\right)  <1$ as the
condition for the fixed point at $1-a$ to be locally stable. Furthermore in
this case also the fixed point at $\frac{1}{\beta}\left(  \frac{1}{2}%
\gamma-a\beta-\frac{1}{2}\sqrt{-4a\beta\gamma+\gamma^{2}}\right)  $ becomes a
local attractor. This is for instance the case when $a$ becomes very small and
$\beta>\gamma$ - being back essentially in the situation of the previous
section. In case when $\frac{1}{\beta}\left(  \frac{1}{2}\gamma-a\beta
+\frac{1}{2}\sqrt{-4a\beta\gamma+\gamma^{2}}\right)  >1$ it is easy to show
that the fixed point at $\frac{1}{\beta}\left(  \frac{1}{2}\gamma-a\beta
-\frac{1}{2}\sqrt{-4a\beta\gamma+\gamma^{2}}\right)  $ becomes a global
attractor (to see this just note that the derivative at $q_{t}=0$ is always
positive for the relevant parameter intervals). The above considerations show
that the possible dynamical evolution scenarios are the same for $a=0$ and
$a\neq0.$ But there is a very strong influence of $a$ on the parameter regimes
of $\beta$ and $\gamma$ for which one has a corruption collapse. Whereas in
case $a=0$ one is always in the basin of attraction of zero for $b_{0}$
sufficiently small and $\gamma\neq0$ (in other words $b=1$ is never a global
attractor) one can now have the phenomenon that only the complete saturation
with corruption is stable ( $q=1-a$). As an example lets look at the case
where $\beta=2\gamma$. For $a=0$ there is a fixed point at $b^{\ast}=0.5$ and
hence for an initial infection density $b_{0}<0.5$ the pure mean field
dynamics converges to zero. In the case $a\neq0$ one has for $a>1/8$ only the
stable fixed point $b^{\ast}=a+q^{\ast}=1.$ At $a=1/8$ there is a phase
transition since a new indifferent (slope 1) fixed point at $b^{\ast}=1/4$
emerges. For $a<1/8$ this fixed point bifurcates into two fixed points where
the first one at $b^{\ast}=\frac{1}{4}-\frac{1}{4}\sqrt{1-8a}$ becomes locally
stable with a basin of attraction given by $b_{0}<$ $\frac{1}{4}+\frac{1}%
{4}\sqrt{1-8a}.$

\begin{figure}
[ptb]
\begin{center}
\includegraphics[
height=2.3321in,
width=3.6129in
]%
{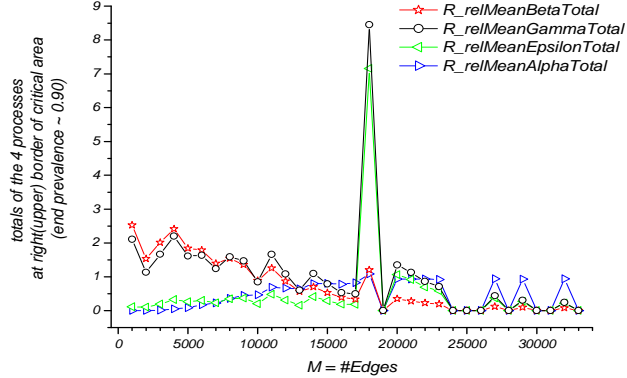}%
\caption{The total number of state changes splitted according to the
$\alpha,\beta,\gamma$ and $\varepsilon$- process for the following parameter
values in a $\mathcal{G}\left(  N,M\right)  $ graph: $N=4000;\alpha
=0.35;\beta=0.08;\gamma=0.04;\varepsilon=0.005;\Delta=5.$}%
\label{200}%
\end{center}
\end{figure}

We close this section by presenting a numerical result showing the different
contributions to the overall infection (end-prevalence) of the local and mean
field processes as a function of the edge density in the random graph space
$\mathcal{G}\left(  N,M\right)$. Fig.\ref{200} gives the accumulated number
of state changes (divided by $N$) caused by the $\alpha,\beta,\gamma$ and
$\varepsilon$- process at initial density values slightly above the critical
one.
Up to an edge density of $2$ (corresponding to a mean degree of $4$) the
$\beta-$ process gives the major contribution to the end prevalence in the
overcritical situation. Parallel to the increase in the edge density increases
the contribution of the $\alpha$- and $\varepsilon$- process (in the
intermediate phase of density between $2$ and $3$ dominated by the $\alpha$-
process) till a sharp peak at edge density $4.5$ where the $\varepsilon-$
process outperforms all the others (at the same time the critical initial
corruption density $b_{0}^{c}$ drops down and becomes almost zero). The peak
is easy to understand since for the chosen parameters we have at an edge
density of $4$ an equality between the recover rate $\gamma$ and the expected
number of new corruptions caused by a single corrupt vertex via the
$\varepsilon-$ process (which is $\mathbb{E}\left(  d\left(  x\right)
\right)  \cdot\varepsilon$). In terms of classical epidemic processes this
corresponds to the case of reproduction number $R_{0}=1.$ Above this value
single initial corrupt vertex is already enough to cause in conjunction with
the mean field process a total infection of the network.

\section{Single run simulation results}

In this section we want to present some simulation results of the corruption
process taking place on some medium size complex networks. Small graph sizes are
interesting as they are typical for communities in highly social structured
populations. As a simple
to generate random graph space with high clustering and power law degree
distribution we have chosen so-called intersection graphs. Intersection graphs
can easily be defined as follows. First one forms random sets from a finite
base set of $N$ elements (random means in this context that the set elements
are chosen uniform i.i.d. from the base set). These sets constitute the
vertices of a random graph. Edges will be defined via the set intersection
property, namely there is an edge between $i$ and $j$ if the associated sets
$A_{i}$ and $A_{j}$ have nonempty intersection. The size (cardinality)
$\left\vert A\right\vert $ of a set $A$ is itself a random variable drawn
i.i.d. from a pre-given probability distribution $\varphi(k).$ To get
interesting graph spaces one furthermore requires $N<\sum\left\vert
A_{i}\right\vert <const\cdot N.$ For theoretical results about the structure
of random intersection graphs see \cite{11}\cite{23}\cite{24}. It is worth
noting that intersection graphs have a high clustering by definition (if an
element is contained in say $k$ sets simultaneously this $k$ sets form a
complete subgraph). Most simulations were done for the case when $\varphi$ is
an asymptotic power law distribution with exponent $3$ or when $\varphi$ is
singular (all sets have the same size). Random intersection graphs have a
multiplicative degree correlation and therefore the critical threshold should
be very low for exponents less than $3$ be the arguments from section 5. Above
that value the form of the degree distribution has only little influence on
the corruption propagation. 

\begin{table}[tbp] \centering
\begin{tabular}
[c]{|l|l|l|}\hline
\textbf{graph characteristic} & \textbf{FP2 (1987-1991)} & \textbf{FP3
(1990-1994)}\\\hline
\# vertices & $4879$ & $7710$\\\hline
\# edges & $57633$ & $93852$\\\hline
mean degree  & $23.624$ & $24.346$\\\hline
maximal degree & $844$ &  $1014$\\\hline
\# vertices wiht degree $>$ $5$ & $3865$ & $6051$\\\hline
size of largest component & $4775$ & $7356$\\\hline
mean \# triangles per vertex & $256.89$ & $418.09$\\\hline
exponent of degree distribution & $2.1$ & $2.4$\\\hline
\end{tabular}%
\caption{Properties of the real networks FP2 and FP3\label{key}}%
\end{table}%

Besides random intersection graphs generated according to some
degree specifications we used also a collection of real collaboration graphs.
These graphs come from a database about research and development projects
funded by the European Community (FP2-3). It's vertices are organizations
involved in European research projects. Two organizations are linked if they
have a joint project (see table II for the main graph characteristics). In total the data base contains about 8000 projects and
13000 participating organizations. In essence the network shows all the main
characteristics that are known from other complex network structures like
scale free degree distribution (with exponent between $2$ and $3$), small
diameter and high clustering and vertex correlation. The initial fraction of
infected individuals was either distributed at random over the vertex set or
clumped together in a sufficiently large ball with a random chosen vertex as center.

\begin{figure}
[ptb]
\begin{center}
\includegraphics[
height=2.1894in,
width=3.5368in
]%
{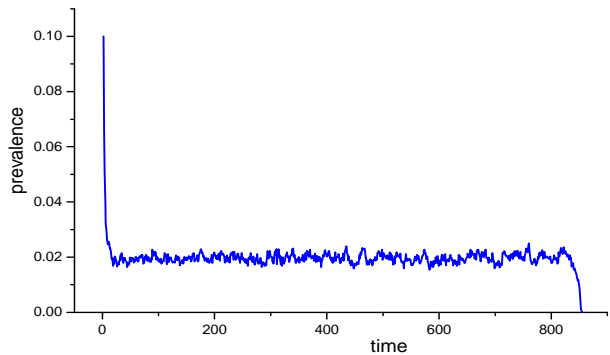}%
\caption{Low, semistable prevalence in a real collaboration network (FP2),
$\Delta$=30 $\varepsilon$=0, $\alpha$=0.99 $\beta$=0.09 $\gamma$=0.545 $b_{0}%
$=0.1 N=4879}%
\label{2}%
\end{center}
\end{figure}

In the following we want to give a small sample of simulations on the just
mentioned graphs and try to discuss its main features. Fig.\ref{2} displays
the prevalence of corruption on the real network FP2. The absolute threshold
value $\Delta=30$ is very high and does not allow for a big outbreak of
corruption. But there is a metastable small community of individuals, highly
linked and almost resistant to the $\gamma-$process. It took more then 800
complete updates till this structure broke down.%

The next figure \ref{3} presents a similar situation on an almost twice as
large real graph (FP3).~
\begin{figure}
[ptb]
\begin{center}
\includegraphics[
height=2.0511in,
width=3.5091in
]%
{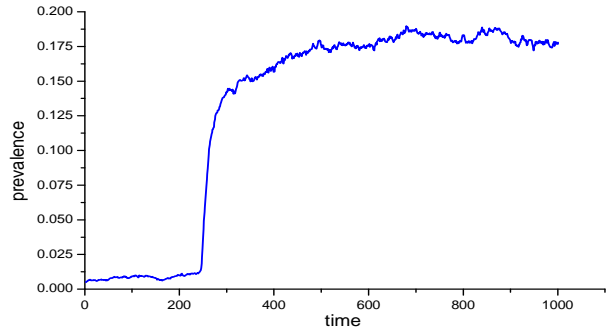}%
\caption{switch from a very low prevalence semi stable state to a low
prevalence state (FP3), $\Delta$=25, $\varepsilon$=0.001, $\alpha$=0.2,
$\beta$=0.04, $\gamma$=0.03, $b_{0}$=0.005, N=7710 }%
\label{3}%
\end{center}
\end{figure}
In contrast to the previous case we have a much smaller $\alpha-$ value and an
only slightly reduced threshold $\Delta$.
\begin{figure}
[ptbptb]
\begin{center}
\includegraphics[
height=2.3001in,
width=3.6691in
]%
{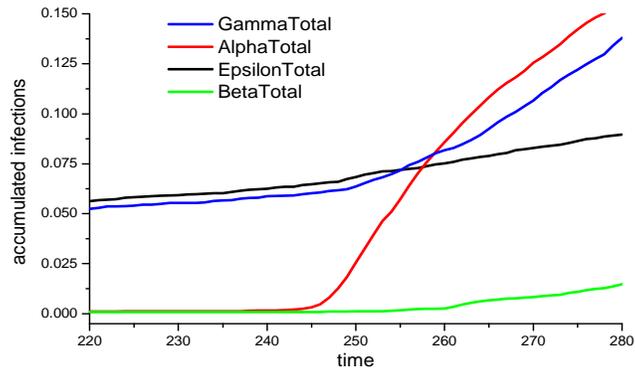}%
\caption{The contribution of the different infection paths for a segment of
the prevalence curve in Fig. \ref{3}}%
\label{4}%
\end{center}
\end{figure}
\begin{figure}
[ptbptbptb]
\begin{center}
\includegraphics[
height=2.2154in,
width=3.7366in
]%
{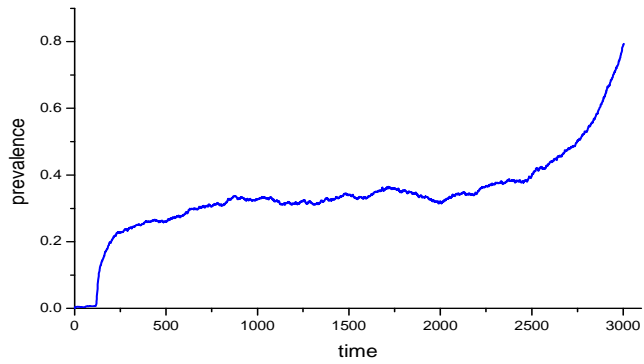}%
\caption{Slow increase of prevalence till collapse (FP3),$\Delta$=20
$\varepsilon$=0.001 $\alpha$=0.2 $\beta$=0.04 $\gamma$=0.03 $b_{0}$=0.005}%
\label{6}%
\end{center}
\end{figure}
\begin{figure}
[ptbptbptbptb]
\begin{center}
\includegraphics[
height=2.2413in,
width=3.5506in
]%
{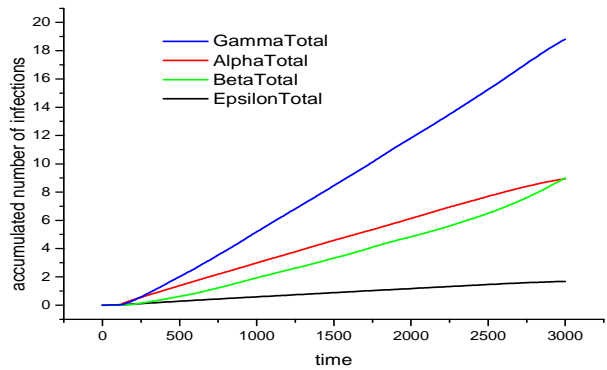}%
\caption{Accumulated infection processes for the situation in Fig.\ref{6}.}%
\label{801}%
\end{center}
\end{figure}
\begin{figure}
[ptbptbptbptbptb]
\begin{center}
\includegraphics[
height=2.2534in,
width=3.6458in
]%
{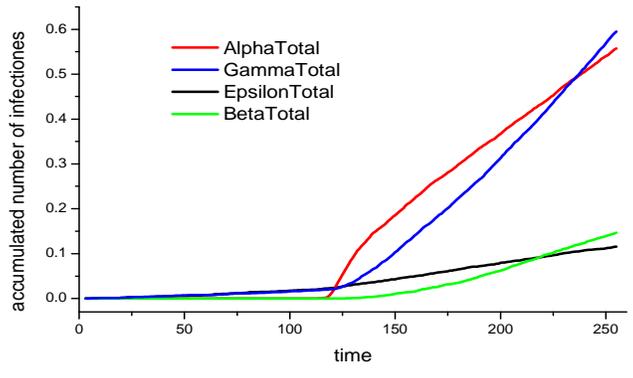}%
\caption{details for the initial phase from fig.\ref{6}. }%
\label{800}%
\end{center}
\end{figure}
\begin{figure}
[ptbptbptbptbptbptb]
\begin{center}
\includegraphics[
height=2.1782in,
width=3.587in
]%
{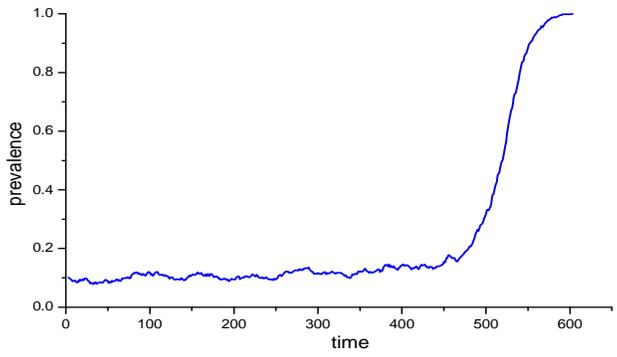}%
\caption{$\Delta$=8 $\varepsilon$=0.005 $\alpha$=0.1 $\beta$=0.09 $\gamma
$=0.045 b$_{0}$=0.1 N=952}%
\label{20}%
\end{center}
\end{figure}
\begin{figure}
[ptbptbptbptbptbptbptb]
\begin{center}
\includegraphics[
height=2.1773in,
width=3.587in
]%
{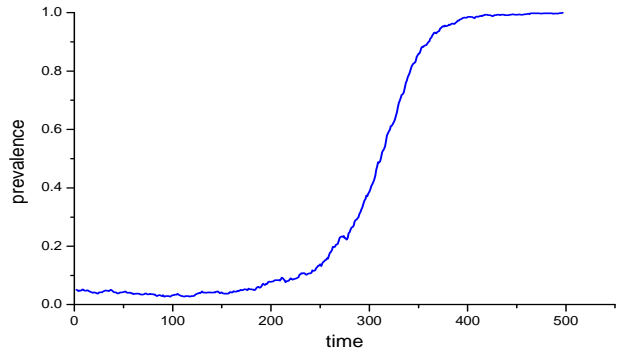}%
\caption{Slow corruption collaps in an artificial net $\Delta$=6 $\varepsilon
$=0.005 $\alpha$=0.1 $\beta$=0.06 $\gamma$=0.05 b$_{0}$=0.05 N=972}%
\label{21}%
\end{center}
\end{figure}
\begin{figure}
[ptbptbptbptbptbptbptbptb]
\begin{center}
\includegraphics[
height=1.9992in,
width=3.3803in
]%
{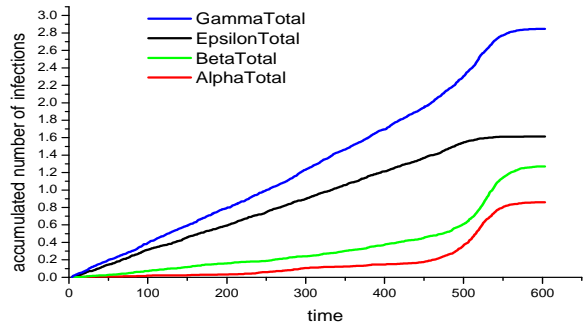}%
\caption{Process splitting for the situation in Fig.\ref{20}}%
\label{22}%
\end{center}
\end{figure}
\begin{figure}
[ptbptbptbptbptbptbptbptbptb]
\begin{center}
\includegraphics[
height=1.9663in,
width=3.3638in
]%
{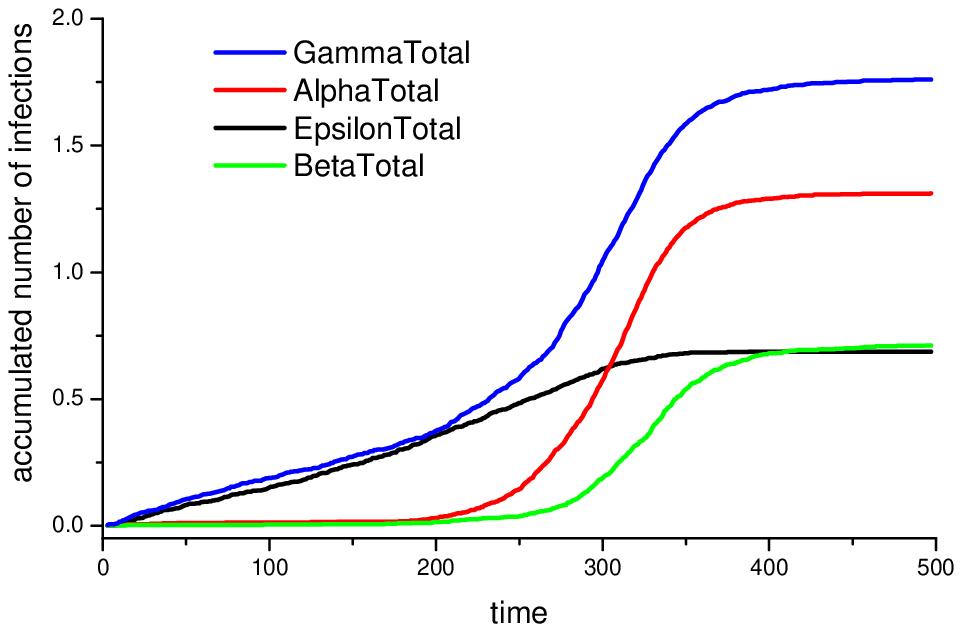}%
\caption{Process splitting for the situation from Fig.\ref{21}}%
\label{23}%
\end{center}
\end{figure}
The network FP3 is extremely high clustered (mean degree $=48.6$, mean
triangle number $=418$ and a total of $7710$ nodes and $187704$ edges) and
stays metastable with a very small corruption cluster for about 200 updates
till it jumps by a factor $10$ to another metastable state. Fig.\ref{4} gives
a more detailed view of the accumulated contributions by the different
processes for a time interval around the jump in prevalence. In the initial
phase the $\varepsilon-$process was dominating the $\beta-$process and vice
versa in the second phase. The next pictures show a situation where after an
initial phase of slow growth a corruption collapse happened. It seems that the
absolute threshold value $\Delta=20$ is well below the critical value where
the system can still stabilize. It is surprising that the system
semi-stabilizes after an initial rapid increase in the prevalence for a rather
long time (Fig.\ref{6}). To a certain extend the results can be explained
along the argumentation in section 7. In Fig.\ref{800} the accumulated
infection processes for the initial phase are shown. Here the $\varepsilon$-
process, although undercritical, causes a redistribution of infection till a
clustered configuration is reached such that the $\alpha$- process can start.
Than the systems stays in almost complete balance till the $\beta-$ process
(which is slow in all our examples) wins (Fig.\ref{801}). Note the difference
to Fig.\ref{4}, where the $\beta-$ process never really contributes to the
infection. Finally we show two simulations for a sample of a random set graph
model with about 1000 vertices (Fig.\ref{20} and Fig.\ref{21}). Although both
prevalence curves look similar there is a clear difference in the process
fine-structure (Fig.\ref{22} and Fig.\ref{23}). In the first instance the
$\varepsilon-$ and $\beta-$ process are causing the collapse whereas in the
second case the $\alpha-$ process in conjunction with the $\varepsilon-$
process is the main booster. 

The few examples of single simulation runs given in this section show already,
that there are many different routes to obtain high prevalence in corruption
typically interrupted by long phases of metastability. Similar to other
complex systems with hidden phase transitions (e.g. the climate) there can be
an unnoticed small accumulation of infection till a critical density- a point
of no return- of corruption is reached from which on an almost complete
saturation of the society (or a corresponding subsystem) by corruption
becomes the normality.

\section{Epidemic control}

One of the basic question in classical epidemics as well as in corruption
dynamics is: what can be done to slow down the "infection" propagation or
prevalence. Knowing the different phase transitions and their dependency on
structure properties and social parameters is of great help in designing
proper prevention scenarios. In the following we will try to relate some of
the findings from our model to what is considered by practicians as useful in
corruption reduction. First we would like to emphasize again that the present
model deals in a rather abstract way with the propagation of mental
willingness to be corrupt and not so much with realized corruption which
always requires a specific environment and additional structural assumptions.
Hence concerning corruption control, we only will be able to support certain
prevention scenarios in the sense, that they go into the right direction and
that there effect is strong or weak but without being able to make
quantitative statements.

The model presented in this paper contains, besides structural parameters for
the underlying network, $5$ relevant parameters: $\alpha$- characterizing the
strength of the local threshold process, $\beta$- characterizing the strength
of the mean field attraction on becoming corrupt, $\gamma$- the strength of
the "society strikes back" term, $\varepsilon$- the strength of the classical
epidemic process (assumed to be very small) and $\Delta$- the height of the
local threshold. Three of the parameters- $\alpha,\beta$ and $\varepsilon$-
are positively correlated to the spread of corruption whereas $2$ parameters-
$\Delta$ and $\gamma$- are negatively correlated. As is well known from
classical epidemic control for infectious diseases it is very hard if not
impossible to change basic social parameters in a short time. This can only be
achieved in a long running educational process. Therefore not much can be done
in avoiding high clustering in certain relevant areas of the society in order
to prevent the emergence of highly connected corruption nets.

As the name already indicates, Transparency International favours as an
effective tool to decrease corruption especially the increase of transparency
in all forms of administrative decision making as well as transparency in
financial affairs of socially exposed persons, institutions and companies. The
effect of an increase of transparency translates into our model as an increase
of the value of $\Delta$ and a decrease of the values of $\beta,\alpha$ and
$\varepsilon$. Strengthening of justice, police and similar instruments to
fight and uncover corruption has again the effect of lowering $\beta$ (via
increase of fear) but may also increase the value of $\gamma$ (uncovering
rate). Since an increase of $\gamma$ above the value of $\beta$ and $\alpha$
would perhaps require a total police state, it is illusionary to overcome
corruption just by means of law, justice and police. Besides necessary long
term educational efforts in school and public to strengthen the moral
resistance against corruption (increase of $\Delta$ and decrease of $\alpha$)
it seems a good strategy to make administrative and political decision
hierarchies as independent and decentralized as possible to avoid high clustering.

We would like to end these short remarks by a few comments on the role of hubs
- the very high degree vertices typically present in scale free graphs - in
corruption dynamics. While a priori not especially well suited to transmit
corruption via the $\alpha-$ process due to the local tree like structure
around the hubs (compared with low degree vertices) they nevertheless are more
often exposed to corruption and have therefore a higher probability to get
corrupt. If the hub density is sufficiently high (as is the case for scale-free degree distributions
with exponent $\lambda<3$) and the degree correlation is
stronger than additive many vertices are linked to the hubs via social
dependencies and in turn also can get corrupt. Furthermore they may play a
fatal role in increasing the weighted corruption density relevant for the mean
field process as was explained at the end of section 7. The described
situation is probably typical for strongly hierarchically organized countries
or regional substructures e.g. systems with a dictatorial or monarchical
tendency. In such societies a high prevalence of corruption seems almost
unavoidable since the threshold $b_{0}^{c}$ is close to zero. For democratic
societies it seems therefore wise, to watch the behavior of hubs- whatever
their social interpretation might be- more intensively than the "normal" part
of the society.

\section{Summary and perspectives}

In this article we have presented a first study of the spread of corruption on
scale free and highly clustered networks. One of the main observations so far
is the strong dependence of the asymptotic dynamics on the initial number of
corrupt individuals. This holds as well for the mean field process as for the
local dynamics. Second there is a fatal resonance effect between global and
local dynamics lowering dramatically the critical density of initial
infection. As expected there is a positive correlation between clustering and
spread of corruption respectively the critical initial density. Scale-freenes
seems to play an important role for the corruption process for distributions
with small exponent ($\lambda<3$) and multiplicative degree correlation due to
the high prevalence of infected hubs and the strong linkage of medium and low
degree vertices to them. For higher exponents the dynamics is rather
insensitive to the degree distribution. The strength of the degree correlation
(from weak - additive till strong - multiplicative or even higher powers) in
networks of social acquaintances seems to be related to the political and
institutional structure of a society which favours liberal organization forms
as being less vulnerable to corruption.

There is a whole bunch of natural continuations or generalizations which have
to be investigated next. Clearly a deeper understanding of the pure $\alpha-$
process and its phase transitions is necessary. The mathematical problem is
already highly nontrivial on trees. The following short list gives a selection
of natural generalizations and refinements:
\begin{description}
\item{--} quenched disorder in all parameters
\item{--} inclusion of geographical or regional structure into the network
\item{--} inclusion of administrative or political substructures in which corruption
typically will be realized
\item{--} evolving networks
\item{--} interaction between the corruption process and the network structure
\item{--} more heterogeneity in the social networks e.g. by incorporating family like
structures or social profiles
\item{--} weighted networks
\item{--} refined transition rules e.g. asymmetry between infecting and getting infected
\item{--} different kinds and strength of corruption and their interplay
\item{--} economic impacts in a virtual population.
\end{description}

Besides the specific context of corruption dynamics there is a multitude of
topics where the model presented in this paper could easily be adopted to.
This includes so different themes as political opinion formation, social
disorder processes, strategies for advertisement, doping usage, the spread of
prejudices, migration dynamics, global terrorist networks and innovation
processes. In all these examples one has a local and global dynamics very
similar to the one described here. Of course there are differences. For
instance in many mind formation problems the state space of individuals is
rather complex and the local dynamics allows for many transitions not just 0-1
as in the corruption model. Furthermore aging phenomena and limits of
resources could be included. But besides this addition of structure and
complexity and the various interpretations there remains a good part of the
findings of this work to be true. There will be phase transitions in the
initial density of certain properties and there can be resonance effects
between the nonlinear global and local dynamics - both making the prediction
of future difficult and challenging.

\bigskip
\noindent
{\large{\bf Acknowledgements}}
We would like to thank for the support of the Volkswagen Foundation, the
DFG-Research Group 399 "Spectral Analysis, Asymptotic Distributions and
Stochastic Dynamics" and the Austrian Research Center Seibersdorf for
providing us with the data set of EU-funded research projects. P. Martin
especially thanks Klaus Geppert, Detlef Leenen, Christian Pestalozza and
Albrecht Randelzhofer for stimulating discussions.

\end{document}